\documentstyle[11pt]{article}
\def\setfonts{%
  \font\frbig=eufm10 scaled\magstephalf
  \font\frscr=eufm8
  \font\frscrscr=eufm8
  \newfam\frfam
  \textfont\frfam=\frbig
  \scriptfont\frfam=\frscr
  \scriptscriptfont\frfam=\frscrscr
  \def\fr{\fam\frfam}

  \font\openbig=msbm10 scaled\magstephalf
  \font\openscr=msbm8 
  \font\openscrscr=msbm8
  \newfam\openfam
  \textfont\openfam=\openbig
  \scriptfont\openfam=\openscr
  \scriptscriptfont\openfam=\openscrscr
  \def\open{\fam\openfam}

  \font\ssfbig=cmss10 scaled\magstephalf
  \font\ssfscr=cmss8 
  \font\ssfscrscr=cmss8
  \newfam\ssffam
  \textfont\ssffam=\ssfbig
  \scriptfont\ssffam=\ssfscr
  \scriptscriptfont\ssffam=\ssfscrscr
  \def\ssf{\fam\ssffam}
  }

\makeatletter
\@addtoreset{equation}{section}
\newdimen\normalarrayskip
\newdimen\minarrayskip
\normalarrayskip\baselineskip
\minarrayskip\jot
\newif\ifold \oldtrue \def\new{\oldfalse}
\def\arraymode{\ifold\relax\else\displaystyle\fi}

\def\@arrayskip{\ifold\baselineskip\z@\lineskip\z@
  \else
  \baselineskip\minarrayskip\lineskip2\minarrayskip\fi}
\def\@arrayclassz{\ifcase \@lastchclass \@acolampacol \or
  \@ampacol \or \or \or \@addamp \or
  \@acolampacol \or \@firstampfalse \@acol \fi
  \edef\@preamble{\@preamble
    \ifcase \@chnum
    \hfil$\relax\arraymode\@sharp$\hfil
    \or $\relax\arraymode\@sharp$\hfil
    \or \hfil$\relax\arraymode\@sharp$\fi}}
\def\@array[#1]#2{\setbox\@arstrutbox=\hbox{\vrule
    height\arraystretch \ht\strutbox
    depth\arraystretch \dp\strutbox
    width\z@}\@mkpream{#2}\edef\@preamble{\halign \noexpand\@halignto
    \bgroup \tabskip\z@ \@arstrut \@preamble \tabskip\z@ \cr}%
  \let\@startpbox\@@startpbox \let\@endpbox\@@endpbox
  \if #1t\vtop \else \if#1b\vbox \else \vcenter \fi\fi
  \bgroup \let\par\relax
  \let\@sharp##\let\protect\relax
  \@arrayskip\@preamble}

\makeatother

\setfonts

\def\req#1{(\ref{#1})}
\def\commut#1#2{\left[{#1},\,{#2}\right]}

\def\frac#1#2{\mathchoice{%
    {\textstyle{{#1}\over{#2}}}}{{#1\over#2}}{{#1\over#2}}{{#1\over#2}}}

\def\half{\frac{1}{2}}

\newtheorem{lemma}{Lemma}[section]

\newtheorem{thm}[lemma]{Theorem}

\newtheorem{dfn}[lemma]{Definition}

{\par\medskip}

\newenvironment{prf}{%
  \noindent{\sc Proof.} \ }%
{\noindent$\Box$\par\medskip} 

\def\Ham{{\open H}}

\def\cA{{\cal A}}

\def\cF{{\cal F}}
\def\cG{{\cal G}}
\def\cH{{\cal H}}
\def\cI{{\cal I}}
\def\cK{{\cal K}}
\def\cL{{\cal L}}
\def\cM{{\cal M}}

\def\cO{{\cal O}}

\def\cR{{\cal R}}
\def\cS{{\cal S}}

\def\PRD{Phys.\ Rev.\ D}
\def\NPB{Nucl.\ Phys.\ B}
\def\PLB{Phys.\ Lett.\ B}
\def\MPLA{Mod.\ Phys.\ Lett.\ A}
\def\CMP{Commun.\ Math.\ Phys.~~}
\def\IJMPA{Int.\ J.\ Mod.\ Phys.\ A}
\def\JMP{J.\ Math.\ Phys.~~}

\newcommand{\abrkt}[2]{({#1}\,,\,{#2})}
\newcommand{\pbrkt}[2]{\{{#1}\,,\,{#2}\}}
\newcommand{\vect}{{\bf Q\,}}
\newcommand{\vectorf}[1]{{\bf #1\,}}
\newcommand{\func}[1]{{{\open C}_{#1}}}

\newcommand{\nullm}[1]{{\cal Z}_{\scriptscriptstyle #1}}
\newcommand{\ideal}[1]{{\open I}_{#1}}
\newcommand{\dr}[1]{\frac{\mathop{\d}\limits^{\rightarrow}}{\d #1}}

\newcommand{\bivect}[1]{{{#1}}}
\newcommand{\twoform}[1]{{\widehat{#1}}}
\newcommand{\Lie}[2]{{{{\pounds}_{#1}}\,{#2}}}

\newcommand{\Ker}{\mathop{{\ssf Ker}}}
\def\Im{\mathop{{\ssf Im}}}
\newcommand{\LG}{\cL}
\def\div#1{\mathop{{\rm div}_{#1}}}
\newcommand{\restrict}[2]{ #1 
  \mbox{\,$\mbox{\rule[-6pt]{0.4pt}{14pt}}_{\,#2} $} }

\def\tilde{\widetilde}

\newcommand\ax{x^*}

\newcommand{\relstack}[2]{\mathop{\rm #1}\limits_{\scriptscriptstyle
    #2}}

\def\d{\partial}

\newcommand{\e}[1]{\epsilon(#1)}
\def\rank#1{{\ssf rank}\left(#1\right)}

\newcommand\manifold{{\sf QP}-manifold}
\newcommand\QP{{\sf QP}}
\newcommand\pmanifold{proper {\sf QP}-manifold}
\newcommand\bv{{\rm BV}-quantization}
\newcommand\m{{\fr m}}
\newcommand\g{{\fr g}}
\newcommand\M{{\fr M}}

\begin{document}
\thispagestyle{empty}
\hfuzz=1.5pt

\addtolength{\baselineskip}{1pt}

\begin{flushright}
  NBI-HE-98-10\\
  {\tt hep-th/9804156}
\end{flushright}

\begin{center}
  {\Large{\sc Gauge Symmetries of the Master Action}}\\[16pt]
  {\large M.~A.~Grigoriev$\,{}^2$, A.~M.~Semikhatov$\,{}^{1,2}$ and
    I.~Yu.~Tipunin$\,{}^2$}\\[12pt]

  \parbox{.85\textwidth}{\small\sl $^1$\,Niels Bohr Institute,
    Blegdamsvej 17, DK-2100,
    Copenhagen\\[6pt]
    $^2$\,Lebedev Physics Institute, Russian Academy of Sciences, 53
    Leninski prosp., Moscow 117924}

  \bigskip
  
  \parbox{.9\textwidth}{\footnotesize We study the geometry of the
    Lagrangian Batalin--Vilkovisky theory on an antisymplectic
    manifold.  We show that gauge symmetries of the BV-theory are
    essentially the symmetries of an {\it even symplectic\/} structure
    on the stationary surface of the master action.}
\end{center}

\section{Introduction}
In this paper, we investigate gauge symmetries in the Lagrangian
Batalin--Vilkovisky (BV) formalism~\cite{[BV],[BV2]}, which is the
most universal approach to the quantization of general gauge theories.
The version of the \bv{} in which the coordinates are not explicitly
separated into fields and antifields is known as the covariant
approach~\cite{[BT],[ASS0],[HZ],[SZ2],[ASS3],[KN],[Bering]}.  The
partition function is then given by a path integral of the exponential
of the master action over the gauge-fixing surface~$\cL$, which is a
Lagrangian submanifold of the odd-symplectic manifold~$\cM$.  The
gauge independence is realized as the independence from the choice
of~$\cL$ and is ensured by the master \hbox{equation imposed on the
  master action.}

While the gauge symmetries of the original action are no longer
explicitly present in the formalism, the covariant formulation itself
has its own ``gauge'' transformations.  Each observable (a BRST-closed
function) determines a gauge symmetry.  Studied in~\cite{[SZ]} were
the gauge symmetries corresponding to the trivial observables
(BRST-exact functions).  It was shown there that the space of
functions modulo the BRST-exact ones, called {\it the space of gauge
  parameters\/}~\cite{[SZ]}, is endowed with the structure of a Lie
algebra, which is induced by the Lie algebra structure on the space of
BRST-trivial gauge symmetries.

In this paper, we study symmetries of the BV formalism using the
geometrical setting provided by viewing the BV data as a
\manifold~\cite{[ASS3],[ASS2]}. These are supermanifolds equipped with
an antisymplectic structure (the {\sf P}-structure) and an odd
nilpotent Hamiltonian vector field (the {\sf Q}-structure); in the BV
setting, the latter is given by the antibracket with the master
action.  An important characteristics of the {\sf Q}-structure is the
{\it zero locus\/} $\nullm\vect$ of the odd vector field.  The most
interesting case in applications is where $\nullm\vect$ is a smooth
$(n|N-n)$-dimensional submanifold (assuming the antisymplectic
supermanifold $\cM$ to be $(N|N)$-dimensional).  We call such
\manifold s the \pmanifold s; then the \QP-structure induces {a
  symplectic structure on the zero locus of~$\vect$
  (see~\req{Poissonbracket}).}\footnote{The existence of a symplectic
  structure on the zero locus can also be inferred from~\cite{[ASS2]};
  the Poisson bracket on $(~,~)$-Lagrangian submanifolds was described
  in~\cite{[DN]}; see also~\cite{[BMS]}.}

It turns out that symmetries of the BV ``master system'' are to a
considerable degree determined by Hamiltonian vector fields on the
stationary surface of the master action (which are Hamiltonian with
respect to the {\it Poisson bracket\/}).  We will explicitly define a
nondegenerate Poisson bracket on the quotient algebra of all smooth
functions modulo the functions vanishing on~$\nullm\vect$; this
generalises the bilinear operation of~\cite{[SZ]}, which was not a
Poisson bracket since it failed to satisfy the Leibnitz rule (in fact,
it was defined on the space that is not an algebra under the
associative multiplication). We show that each symmetry of a
\pmanifold{}---i.e., a vector field preserving the \QP-structure---can
be restricted to $\nullm\vect$ and, moreover, this restriction is a
locally Hamiltonian vector field with respect to the Poisson bracket
on~$\nullm\vect$.
Conversely, locally Hamiltonian vector fields on $\nullm\vect$ can be
lifted to vector fields on $\cM$, into symmetries of the \pmanifold.
At the same time, the {\it globally\/} Hamiltonian vector fields on
$\nullm\vect$ lift to BRST-trivial symmetries. In this way, we obtain
a ``translation table'' between the objects pertaining to the
antisymplectic geometry on $\cM$ and to those of the symplectic
geometry on~$\nullm\vect$.

We further select the {\it on-shell symmetries}, i.e., the symmetries
modulo those vanishing on the stationary surface.  We show that the
Lie algebra $\Ham_{\nullm\vect}$ of on-shell gauge symmetries is
isomorphic to the Lie algebra of locally Hamiltonian vector fields on
the stationary surface~$\nullm\vect$.

The Lie algebras of on-shell gauge symmetries ($\Ham_{\nullm\vect}$),
of gauge parameters~\cite{[SZ]} ($\tilde\cO^c_{\rm triv}$), and of
gauge symmetries of the master action ($\cO^c$), as well as their
quantum counterparts, are related to each other as shown
in~\req{maindiag} and to the cohomology, as shown in~\req{exact-1},
\req{exact-2}, and~\req{exact-3}.

We consider two examples of the general construction.  We explicitly
calculate the Lie algebras $\cO^c$ and $\Ham_{\nullm\vect}$ in the
{\it abelianized\/}~\cite{[B-abel]} gauge theory. It is not difficult
to see then that the algebra of gauge symmetries of the original
(``bare'') classical theory is embedded into the Lie algebra~$\cO^c$
as a subalgebra in such a way that the algebra of on-shell gauge
symmetries of the original theory is embedded into the Lie
algebra~$\Ham_{\nullm\vect}$.  As another example, we consider the
theory with the vanishing action on a Lie group. Not surprisingly, the
Poisson structure on the ``stationary surface'' is then related to the
Kirillov bracket~\cite{[Kir]} on the coalgebra.

\smallskip

In section \ref{geometry}, we study the geometry of \manifold s.  In
section~\ref{def:gaugesymm}, we give a short reminder on the
BV-quantization prescription and then study the quantum and classical
gauge symmetries.  In section~\ref{Examples}, we demonstrate the main
points of our construction in two characteristic examples.

\section{Geometry of \pmanifold s\label{geometry}}
In this section, we study the geometry of \manifold s and define
\pmanifold s, which are needed for applications to the \bv{}. Then we
show that the zero locus $\nullm\vect$ of the vector field $\vect$ on
a \pmanifold{} is a symplectic manifold.  Moreover, the vector fields
that are {\it symmetries\/} of a \pmanifold{} correspond to locally
Hamiltonian vector fields on $\nullm\vect$; the {\it BRST-trivial\/}
symmetries then correspond to globally Hamiltonian vector fields.

\subsection{A Poisson structure}
Let $\cM$ be an $(N|N)$-dimensional supermanifold, and let $\func\cM$
denote the algebra of smooth functions on $\cM$.  Let
$\abrkt\cdot\cdot\,:\,\func\cM\times\func\cM\rightarrow\func\cM$ be an
antibracket on~$\cM$. It satisfies
\begin{eqnarray}
  \e{\abrkt{F}{G}}&=&\e F+\e G+1\,,\\
  \abrkt{F}{G}&=&-(-1)^{(\e F+1)(\e G+1)}\abrkt{G}{F}\,,\\
  \abrkt{F}{GH}&=&\abrkt{F}{G}H+(-1)^{\e G(\e F+1)}G\abrkt{F}{H}\,,\\
  0&=&\relstack{cycle}{F,G,H}(-1)^{(\e F+1)(\e H+1)}
  \abrkt{F}{\abrkt{G}{H}}\,.
\end{eqnarray}
In a local coordinate system $\Gamma^A$, $A=1,\ldots,2N$, we have the
matrix $E^{AB}=\abrkt{\Gamma^A}{\Gamma^B}$ that defines a
bivector~$\bivect E$ such that $\abrkt{F}{G}=\bivect E(dF,dG)$.  We
assume the antibracket to be nondegenerate.

Consider an odd vector field $\vect\,:\,\func\cM\to\func\cM$ on~$\cM$
satisfying the following conditions:
\begin{enumerate}\addtolength{\parskip}{-6pt}
\item $\vect$ preserves the antibracket, i.e., $\Lie{\vect}{\bivect
    E}=0$, where $\Lie{\vect}{}$ is the Lie derivative along~$\vect$;
  equivalently, $\vect$ differentiates the antibracket:
  \begin{equation}
    \vect\abrkt{F}{G}
    -\abrkt{\vect F}{G}-(-1)^{\e F+1}\abrkt{F}{\vect G}=0\,,
    \quad F,G\in\func\cM\,;
    \label{Leibnitz}
  \end{equation}
\item $\vect$ is nilpotent: $\vect(\vect F)=0\,,~F\in\func\cM\,,\
  \Longleftrightarrow\ \commut\vect\vect=0$.
\end{enumerate}
\begin{dfn}[\cite{[ASS3],[ASS2]}] \label{def:QP}
  A supermanifold $\cM$ equipped with a nondegenerate antibracket
  $\abrkt\cdot\cdot$ and an odd nilpotent vector field $\vect$ that
  satisfies condition~\req{Leibnitz} is called the \manifold.
\end{dfn}

The main object of our analysis is the set $\nullm\vect$ of zeroes of
$\vect$, i.e. the set defined by equations~$Q^A=0$,
where~$\vect=Q^A\d_A$ in a local coordinate system $\Gamma^A$.  We
assume $\nullm\vect$ to be a submanifold in~$\cM$.  Denote by
$\ideal{\nullm\vect}\subset\func\cM$ the ideal of all smooth functions
vanishing on~$\nullm\vect$. We also assume the `regularity condition',
i.e. that $\vect$-exact functions generate the ideal
$\ideal{\nullm\vect}$, which means that any function $f \in
\ideal{\nullm\vect}$ admits a representation $f=(\vect h)\,g$ with some
$g,h \in \func\cM$.  The quotient
$\func{\nullm\vect}=\func\cM/\ideal{\nullm\vect}$ is the algebra of
smooth functions on the sub\-mani\-fold~$\nullm\vect$.  Obviously,
$\vect\func\cM\subset\ideal{\nullm\vect}$.

The additional requirement imposed on $\vect$ is that its local
cohomology (that is, the cohomology evaluated in a sufficiently small
neighbourhood of a point) is trivial (constants only) at every point
$p\in\nullm\vect$. In local coordinates, then, the nilpotent operator
$\restrict{\d Q^A/\d\Gamma^B}{Q^A=0}$ has the vanishing cohomology on
the tangent space to every $p\in\nullm\vect$~\cite{[ASS3],[ASS2]}.
This in turn implies the condition from~\cite{[BV]}:
\begin{equation}
    \left.\rank{\d Q^A\over\d\Gamma^B}\right|_{Q^A=0}=N\,.
  \label{rankcondit}
\end{equation}
In particular, it follows from \req{rankcondit} that $\nullm\vect$ is
$(n|N-n)$-dimensional submanifold.
\begin{dfn} \label{def:PQP}
  A \pmanifold{} is a \manifold{} on which the local cohomology of
  $\,\vect$ is trivial at every point from $\nullm\vect$.
\end{dfn}
\begin{lemma}\label{closeideal}
  The submanifold $\nullm\vect$ of a \pmanifold{} is Lagrangian with
  respect to the antibracket.  In particular, the ideal
  $\ideal{\nullm\vect}$ is closed under the antibracket{\rm:}
  \begin{equation}
    \abrkt{\ideal{\nullm\vect}}{\ideal{\nullm\vect}}
    \subset
    {\ideal{\nullm\vect}}\,.
  \end{equation}
\end{lemma}
\begin{prf}
  Since $\vect$-exact functions {\it generate\/} the
  ideal~$\ideal{\nullm\vect}$, any function from $\ideal{\nullm\vect}$
  can be represented as the product $(\vect f)\,h$ with some
  $h\in\func\cM$. Thus, it suffices to check that the antibracket of
  $\vect$-exact functions is $\vect$-exact, which is obvious in view
  of \ $\abrkt{\vect g}{\vect h}=\vect\abrkt{g}{\vect h}$.
\end{prf}

In fact the submanifold $\nullm\vect$ is endowed with a natural
Poisson structure. This is given by a construction of the type of
those used, with some variations, in~\cite{[SZ],[DN],[BMS],[Ners]},
namely
\begin{equation}\label{Poissonbracket}
  \pbrkt{F}{G}=\abrkt{F}{\vect G}\,,\quad F,G\in\func\cM\,.
\end{equation}
We interpret this structure as a bilinear mapping on the quotient
algebra $\func{\nullm\vect}$. Functions from $\func\cM$ considered
modulo $\ideal{\nullm\vect}$ represent functions on~$\nullm\vect$.  We
then have
\begin{thm}\label{thm:symplectic}
  For any \pmanifold,
  \begin{enumerate}\addtolength{\parskip}{-4pt}
  \item Equation~\req{Poissonbracket} defines a Poisson bracket
    $\pbrkt\cdot\cdot\,:\,\func{\nullm\vect}\times\func{\nullm\vect}
    \rightarrow\func{\nullm\vect}$ on the submanifold
    $\nullm\vect${\rm};
  \item moreover, the Poisson bracket $\pbrkt\cdot\cdot$ is
    nondegenerate (thus, $\nullm\vect$ is symplectic).
  \end{enumerate}
\end{thm}\pagebreak[3]
\begin{prf}
  First of all, we must prove that definition~\req{Poissonbracket}
  does not depend on the choice of representatives of the equivalence
  classes, i.e., $\pbrkt{F+f}{G+g}-\pbrkt{F}{G}\in\ideal{\nullm\vect}$
  whenever~$f,g\in\ideal{\nullm\vect}$.  Since $\vect$ differentiates
  the antibracket, we can check that
$$
    \pbrkt{F+f}{G+g}-\pbrkt{F}{G}=
    (-1)^{\e F}\abrkt{\vect F}{g}+(-1)^{\e F+1}\vect\abrkt{F}{g}
    +\abrkt{f}{\vect (G+g)}\,.
$$
  The first and the third terms belong to~$\ideal{\nullm\vect}$ by
  Lemma~\ref{closeideal} and the second term is
  in~$\ideal{\nullm\vect}$ because it is $\vect$-exact. Thus
  \req{Poissonbracket} defines a mapping
  $\pbrkt\cdot\cdot\,:\,\func{\nullm\vect}
  \times\func{\nullm\vect}\rightarrow\func{\nullm\vect}$.  It is
  antisymmetric because
$$
\new
    \begin{array}{rcl}
      \pbrkt{F}{G}+(-1)^{\e F \e G}\pbrkt{G}{F}&=&
      \abrkt{F}{\vect G}+(-1)^{\e F \e G}\abrkt{G}{\vect F}\\
      {}&=&
      (-1)^{\e{F}+1}\vect\abrkt{F}{G}\in\ideal{\nullm\vect}\,,
    \end{array}
$$
  where we used~\req{Leibnitz} again.  Next, to prove the Leibnitz
  rule, we evaluate
$$
\new
    \begin{array}{l}
      \pbrkt{F}{G H}
      -\pbrkt{F}{G}H
      -(-1)^{\e{F} \e{G}}G\pbrkt{F}{H}\\
      \kern5cm{}=(-1)^{(\e{F}+1)(\e{G}+1)}(\vect G)\abrkt{F}{H}
      +(-1)^{\e{G}}\abrkt{F}{G}(\vect H)\,,
    \end{array}
$$
  which evidently vanishes modulo terms from~$\ideal{\nullm\vect}$.
  Finally, to prove the Jacobi identity we have to show that
  $$
    \relstack{cycle}{F,G,H}\,(-1)^{\e F\e H}\pbrkt{F}{\pbrkt{G}{H}}=
    \relstack{cycle}{F,G,H}\,
    (-1)^{\e F\e H}\abrkt{F}{\vect\abrkt{G}{\vect H}}\equiv0~{\rm mod}~
    \ideal{\vect}
$$
  In view of the Leibnitz rule and the nilpotency condition this
  rewrites, modulo terms from~$\ideal{\nullm\vect}$, as
 $$
    \relstack{cycle}{\vect F,\,G,\,\vect H}
    (-1)^{(\e{\vect F}+1)(\e{\vect H}+1)}
    \abrkt{\vect F}{\abrkt{G}{\vect H}}\,,
$$
  which vanishes by virtue of the Jacobi identity for the antibracket.

  To prove that the Poisson bracket is nondegenerate on~$\nullm\vect$,
  we recall a standard fact from symplectic geometry, namely that in
  some neighborhood of~$\nullm\vect$ there exists a coordinate
  system~$x^i$, $\xi_i$ such that the antibracket takes the canonical
  form~$\abrkt{x^i}{\xi_j}=\delta^i_j$ and~$\nullm\vect$ is determined
  by~$\xi_i=0$. Locally, the vector field~$\vect$ can be written in
  the form~$\vect=\abrkt{S}{\;\cdot\;}$ with some
  function~$S\in\func\cM$ (since a vector field preserving a
  nondegenerate (anti)bracket is locally Hamiltonian).
  Expanding~$S$~as
  $$
  S=S_0(x)+\xi_iS^i(x)+\xi_iS^{ij}(x)\xi_j
  +\xi_i\xi_j\xi_kS^{ijk}(x)+\ldots\,,
  $$
  we see that~$S_0(x)={\rm const}$ and $S^i(x)=0$, because
  $\vect=\abrkt{S}{\cdot~}$ vanishes as $\xi=0$.  Then
  condition~\req{rankcondit} means that the matrix~$S^{ij}(x)$ is
  non-degenerate at each point of~$\nullm\vect$.  On the other hand,
  $S^{ij}(x)|_{\xi=0}=-\half\pbrkt{x^i}{x^j}|_{\xi=0}$, which shows
  the theorem.
\end{prf}

Note that the symbol $\pbrkt\cdot\cdot$ is used for the formal
operation~\req{Poissonbracket} on the manifold~$\cM$ and also for the
Poisson bracket on the submanifold~$\nullm\vect$.  We do not introduce
two different symbols and hope that this will not lead to confusion.

\subsection{Symmetries of \manifold s\label{QPsymm:subsec}}
For applications to the BV-quantization in the subsequent sections, we
will need some facts about symmetries of \manifold s.
\begin{dfn}\label{def:PQPsymm}
  A vector field $\vectorf X$ on a \manifold{} $\cM$ is called a
  symmetry of $\cM$ if
  \begin{enumerate}\addtolength{\parskip}{-6pt}
  \item $\vectorf X$ preserves the
    antibracket~$\abrkt\cdot\cdot${\rm:}
    \begin{equation}\label{LeibnitzX}
      \vectorf X\abrkt{F}{G}-
      \abrkt{\vectorf X F}{G}-
      (-1)^{(\e F+1)\e{\vectorf X}}\abrkt{F}{\vectorf X G}=0\,,
      \qquad  F,G\in \func\cM\,,\pagebreak[3]
    \end{equation}

  \item $\vectorf X$ preserves the odd vector field~$\vect${\rm:}~
    $\commut{\vect}{\vectorf X}=0\,$.
  \end{enumerate}
\end{dfn}

Our aim is to demonstrate that symmetries of a \pmanifold{} restrict
to the zero locus of~$\vect$ and to study the properties of these
restrictions. Let us therefore begin with characterising, in the
standard way, those vector fields on $\cM$ that restrict to
$\nullm\vect$:
\begin{lemma}\label{thm:nesenaugh}
  A vector field $\vectorf X$ on a \pmanifold{} $\cM$ restricts
  to~$\nullm\vect$ if and only if
  \begin{equation}\label{commutonconstr}
    \restrict{\commut{\vectorf X}{\vect}}{\nullm\vect}=0\,.
  \end{equation}
\end{lemma}
\begin{prf}
  A vector field restricts to $\nullm\vect$ if and only if it
  preserves the ideal of functions vanishing on $\nullm\vect$:
  ${\vectorf X}\ideal{\nullm\vect}\subset\ideal{\nullm\vect}$. Now,
  $\restrict{\commut{\vectorf X}{\vect}}{\nullm\vect}=0$
  $\Longleftrightarrow$ $[{\vectorf X},\vect]f\in\ideal{\nullm\vect}$
  $\forall f\in\func\cM$, which rewrites as ${\vectorf X}\vect
  f\in\ideal{\nullm\vect}$. Since $\vect$-exact functions generate the
  ideal, we conclude that $\vectorf X \ideal{\nullm\vect} \subset
  \ideal{\nullm\vect}$. The converse is now obvious.
\end{prf}

It follows from this Lemma that any vector field $\vectorf X$ that is
a symmetry of a \pmanifold{} can be restricted to~$\nullm\vect$. We
now recall that the zero locus of~$\vect$ is endowed with a Poisson
structure.
\begin{thm}\label{thm:symm-restrict}
  If a vector field $\vectorf X$ is a symmetry of a \pmanifold{}
  $\cM$, its restriction~$\restrict{\vectorf X}{\nullm\vect}$ to the
  zero locus of $\vect$ preserves the Poisson bracket from
  Theorem~\ref{thm:symplectic}:
  \begin{equation}
    \restrict{\vectorf X}{\nullm\vect}\pbrkt{F}{G}-
    \pbrkt{\restrict{\vectorf X}{\nullm\vect} F}{G}
    -(-1)^{\e F\e{\vectorf X}}
    \pbrkt{F}{\restrict{\vectorf X}{\nullm\vect} G}=0\,,
    \qquad F,G\in\func{\nullm\vect}\,.
  \end{equation}
\end{thm}
\begin{prf}
  Let us choose two representatives~$F,G\in\func\cM$ of the
  equivalence classes of functions on~$\nullm\vect$.  Using the
  properties stated in Definition~\ref{def:PQPsymm}, we have
$$
    \begin{array}{rcl}
      \vectorf X\pbrkt{F}{G}&=&\vectorf X\abrkt{F}{\vect G}=
      \abrkt{\vectorf X F}{\vect G}
      +(-1)^{(\e F+1)\e{\vectorf X}}\abrkt{F}{\vectorf X\vect G}\\
      {}&=&\pbrkt{\vectorf X F}{G}
      +(-1)^{\e F\e{\vectorf X}}\pbrkt{F}{\vectorf X G}.
    \end{array}
$$
\end{prf}
It follows from the nondegeneracy of the Poisson
bracket~\req{Poissonbracket} that any vector field $\vectorf x$ on
$\nullm\vect$ that preserves the Poisson bracket can be written as
$\vectorf x=\pbrkt{H}{\cdot\,}$ with some (locally) defined
function~$H$. We will refer to this as a {\it locally Hamiltonian\/}
vector field.  Every $\vectorf x=\pbrkt{H}{\cdot\,}$ with a globally
defined $H$ will be called {\it globally Hamiltonian}.  It is well
known that globally Hamiltonian vector fields
form an ideal in the Lie algebra of locally Hamiltonian vector fields.

\subsection{Lifts and restrictions of vector fields}
In this section, we are interested in relations between Hamiltonian
vector fields on the symplectic manifold $\nullm\vect$ and the
Hamiltonian vector fields on $\cM$,\footnote{ Thus, whenever we speak
  about Hamiltonian vector fields on $\cM$, these are Hamiltonian with
  respect to the {\it anti\/}bracket, while the Hamiltonian vector
  fields on $\nullm\vect$ are Hamiltonian with respect to the Poisson
  bracket.} in particular, symmetries of the \pmanifold~$\cM$.

To explain why there exists a correspondence between symmetries of
\pmanifold{}~$\cM$ and symmetries of its symplectic submanifold
$\nullm\vect$, we begin with an example~\cite{[Ners]}.  Consider an
$N$-dimensional symplectic manifold~$\cK$ with the symplectic
form~$\twoform\omega$ which defines the nondegenerate Poisson
bracket~$\pbrkt{\cdot}{\cdot}$ (in general, $\cK$ can be a
supermanifold, but we assume for simplicity that it is an even
manifold).  In a local coordinate system $x^i$, we have the invertible
matrix~$\omega^{ij}=\pbrkt{x^i}{x^j}$.  Let $\Pi T^*\cK$ be the
cotangent bundle with the flipped parity.  In the canonical
coordinates $x^i,\;\xi_i$ (with the antibracket
$\abrkt{x^i}{\xi_j}=\delta^i_j$), the manifold $\cK$ can be identified
with the zero section~$\xi_i=0$ of~$\;\Pi T^*\cK$.  The
function~$S=\half\xi_i\omega^{ij}\xi_j$ satisfies~$\abrkt{S}{S}=0$
because~$\omega^{ij}$ is the matrix of a Poisson bracket.  Then the
\hbox{submanifold~$\cK$ is the zero locus of the vector field}
\begin{equation}
  \vect=\abrkt{S}{\cdot\;}=
  \half \xi_i(\dr{x^k}\omega^{ij})\xi_j\dr{\xi_k}
  -\xi_i\omega^{ij}\dr{x^j}\,.
\end{equation}
It is easy to see that $\vect$ meets the conditions of
Definition~\ref{def:QP}.  In addition, condition~\req{rankcondit} is
satisfied because the dimension of $\cK$ is~$N$.  Thus $\Pi T^*\cK$ is
a \manifold{} and in fact a \pmanifold{} (because~$\omega^{ij}$ is
nondegenerate).  With the help of the symplectic form, we can identify
$\Pi T^*\cK$ with the tangent bundle $\Pi T\cK$ (with
$\xi^i=\omega^{ij}\xi_j$ being the coordinates on the fibers). Then we
can rewrite~$\vect$~as
\begin{equation}
  \vect=\xi^i\dr{x^i}\,,
  \label{Qcanonical}
\end{equation}
Upon the identification of functions on $\Pi T\cK$ with differential
forms on $\cK$, \ $\vect$ becomes the De Rham differential
on~$\cK$~\cite{[ASS2]}.

Further, every locally Hamiltonian vector field $\vectorf x$ on~$\cK$
can be lifted to a globally Hamiltonian vector field $\vectorf X$
on~$\Pi T^*\cK$.  Namely, if $\vectorf x=\pbrkt{H}{\cdot\,}$ (where we
allow $H$ to be multivalued), we take $F=(-1)^{\e{H}}\vect(\pi^*H)$,
where $\pi^*$ is the pullback with respect to the canonical
projection~$\pi:\Pi T^*\cK\to\cK$. Then the vector field $\vectorf
X=\abrkt{F}{\cdot\,}$ {\it is well-defined\/} (independent of the
multivaluedness of~$H$) and satisfies
\begin{equation}
  \restrict{\vectorf X}{\cK}=\vectorf x
\end{equation}
and is a symmetry of $\Pi T^*\cK$ in the sense of
Definition~\ref{def:PQPsymm}. Conversely, any symmetry of $\Pi T^*\cK$
determines a locally Hamiltonian vector field on $\cK$ (see Theorem
\ref{thm:symm-restrict}), which is obviously a symmetry of this
symplectic manifold.

\smallskip

The above is a particular case of a more general construction.
Namely, there exists a similar correspondence between symmetries of
\pmanifold{}~$\cM$ and symmetries of its symplectic submanifold
$\nullm\vect$, even though $\cM$ can be an arbitrary \pmanifold{}.

We have seen in theorem \ref{thm:symm-restrict} that every symmetry of
a \pmanifold{} $\cM$ restricts to $\nullm\vect$ as a locally
Hamiltonian vector field.  Consider now the converse problem.  We say
that a symmetry~$\vectorf X$ of \pmanifold~$\cM$ is a lift
from~$\nullm\vect$ of a locally Hamiltonian vector field~$\vectorf x$
if $\vectorf X$ restricts to~$\nullm\vect$ and $\restrict{\vectorf
  X}{\nullm\vect}=\vectorf x$.
\begin{thm}\label{thm:lift}
  Every locally Hamiltonian vector field $\vectorf x$ on $\nullm\vect$
  admits a lift to a symmetry of $\cM$ that is a globally Hamiltonian
  vector field on $\cM$ with a $\vect$-closed Hamiltonian.  If
  $\vectorf x$ is globally Hamiltonian on $\nullm\vect$, it is lifted
  to a globally Hamiltonian vector field with a $\vect$-exact
  Hamiltonian.
\end{thm}
In what follows, symmetries of $\cM$ of the form $\vectorf
X=\abrkt{F}{\;\cdot\;}$ with a $\vect$-exact Hamiltonian $F$ are
called {\it BRST-trivial symmetries}.

\begin{prf} For a (locally) Hamiltonian vector field $\vectorf x$ on
  $\nullm\vect$, the equation $\vectorf x=\pbrkt{H}{\cdot\,}$ can be
  solved for $H$ in a sufficiently small neighbourhood of every
  point~$p\in\nullm\vect$.  Different such solutions can be considered
  as a multivalued Hamiltonian~$H$.  This can be extended to a
  multivalued function $\tilde H$ on $\cM$ that restricts to $H$ on
  $\nullm\vect$ (for example, consider a neighbourhood $U$ of
  $\nullm\vect$ in $\cM$ and identify it with a neighbourhood of the
  zero section of a vector bundle over $\nullm\vect$; if $\tilde h$ is
  the pullback of the multivalued function $H$ to the bundle, we can
  choose a function $\alpha \in \func\cM$ such that
  $\alpha|_{\nullm\vect}=1$ and $\alpha=0$ outside $U$, which yields
  the lifting ${\tilde H}=\alpha{\tilde h}$ of the multivalued
  Hamiltonian~$H$).

  Then consider the function $F=(-1)^{\e{H}}\vect{\tilde H}$ on $\cM$;
  $F$ is a {\it single-valued\/} function that is $\vect$-closed by
  construction but in general is not $\vect$-exact (because its
  $\vect$-primitive is not necessarily single-valued).  Now, let
  $\vectorf X=\abrkt{F}{\cdot\,}$.  For any function ${\tilde
    G}\in\func\cM$, we have
  $$
  \abrkt{F}{\tilde G}\Bigm|_{\nullm\vect} = \abrkt{\tilde
    H}{\vect{\tilde G}}\Bigm|_{\nullm\vect}\,,
  $$
  which coincides with $\pbrkt{\restrict{\tilde H}{\nullm\vect}}
  {\restrict{\tilde G}{\nullm\vect}}= \vectorf x \restrict{\tilde
    G}{\nullm\vect}$ (see~\req{Poissonbracket}).  Thus $\vectorf X$ is
  a lift of $\vectorf x$ to a symmetry~of~$\cM$.
  
  Whenever the Hamiltonian $H$ of $\vectorf x$ on $\nullm\vect$ is
  globally defined on $\nullm\vect$, the function $F$ is obviously
  $\vect$-exact and thus $\vectorf X =\abrkt{F}{\cdot\,}$ is a BRST
  trivial symmetry of $\cM$.
\end{prf}
This theorem can also be seen by noticing that the cohomology of
$\vect$ evaluated on an appropriately chosen neighbourhood $U$ of
$\nullm\vect$ in $\cM$ coincides with the De Rham cohomology
of~$\nullm\vect$~\cite{[ASS2]}.  Namely, one can identify the
neighbourhood $U$ of $\nullm\vect$ with some neighbourhood of the zero
section of~$\Pi T \nullm\vect$.  Then vector field $\vect$ can be
written as $\vect= \xi^i \dr{x^i}$, where $x^i$ and $\xi^i$ are
coordinates on $\nullm\vect$ and on the fibers, respectively.  Thus
$\vect$ coincides with the De Rham differential of $\nullm\vect$ if
one identifies functions on $U$ that are homogeneous in $\xi$ with the
differential forms on $\nullm\vect$.  In particular, every closed but
not exact 1-form $f=dx^if_i$ leads to the function $F=\xi^if_i$ on $U$
that is obviously in the cohomology of~$\vect$.  At the same time, the
1-form $f=dx^if_i$ gives rise to the locally Hamiltonian vector field
$\vectorf x=(-1)^{\e{x^i}\e{f_i}}f_i \omega^{ij}\dr{x^j}$
on~$\nullm\vect$ (where $\omega^{ij}=\pbrkt{x^i}{x^j}$).  Therefore
$\vectorf x$ lifts to the vector field
$(-1)^{\e{F}+1}\abrkt{F}{\cdot\,}$ on $U$ whose Hamiltonian is the
same function $F=\xi^if_i$.

We thus see, in particular, that a locally Hamiltonian vector field
representing the first cohomology of~$\nullm\vect$ corresponds to an
element of the $\vect$-cohomology on~$\cM$.
To complete the section let us single out those
$\abrkt{~\cdot}{\cdot~}$-Hamiltonian vector fields on~$\cM$ that
restrict to $\pbrkt{~\cdot}{\cdot~}$-Hamiltonian vector fields
on~$\nullm\vect$, and describe the full arbitrariness of the lifts of
Hamiltonian vector fields on~$\nullm\vect$ to Hamiltonian vector
fields on~$\cM$. The following is proved by directly generalising the
proof of Theorem~\ref{thm:lift}.
\begin{thm}\label{thm:restrict} {$\mbox{ }$ }
  \begin{enumerate}\addtolength{\parskip}{-6pt}

  \item Let~$\vectorf X=\abrkt{F}{\cdot\,}$ be a globally Hamiltonian
    vector field on a \pmanifold~$\cM$ with the Hamiltonian satisfying
    $\vect F \in \ideal{\nullm\vect}^3$ {\rm(}i.e., $\vect
    F=Q^AQ^BQ^CY_{ABC}$ with some $Y_{ABC}\in\func\cM${\rm)}.  Then
    $\vectorf X$ restricts to a locally Hamiltonian vector field
    on~$\nullm\vect$.
  \item Every locally Hamiltonian vector field $\vectorf x$ on
    $\nullm\vect$ admits a lift to a globally Hamiltonian vector field
    on $\cM$ with the Hamiltonian $F$ satisfying $\vect F=0$.  If
    $\vectorf x$ is globally Hamiltonian on $\nullm\vect$ with the
    Hamiltonian~$H$, the Hamiltonians of all its lifts to $\cM$ are of
    the form
    \begin{equation}
      F=(-1)^{\e{\tilde H}}\vect{\tilde H}+K+{\rm const}\,,
      \quad K\in \ideal{\nullm\vect}^2,
      \label{Hlift}
    \end{equation}
    where $\tilde H$ is any function on~$\cM$ such that
    $\restrict{\tilde H}{\nullm\vect}=H$.
  \end{enumerate}
\end{thm}

\section{Gauge symmetries of the master-action\label{def:gaugesymm}}
We now interpret classical gauge symmetries in the covariant BV
formalism as symmetries of the corresponding \pmanifold.  Using the
results of the previous section, we then show that the Lie algebra of
locally Hamiltonian vector fields on the stationary surface of the
master action coincides with the algebra of {\it on-shell gauge
  symmetries}.  Section~\ref{subsec:BV-reminder} contains a brief
reminder on the BV formalism, so the reader may wish to go directly to
Sec.~\ref{sec:gauge-symmetries}.

\subsection{Batalin--Vilkovisky quantization\label{subsec:BV-reminder}}
The geometrical background of the covariant formulation of the BV
quantization is an $(N|N)$-di\-mensional supermanifold $\cM$ equipped
with a nondegenerate antibracket $\abrkt\cdot\cdot$ and a volume form
$d\mu=\rho d\Gamma$, where $\rho=\rho(\Gamma)$ is a density (and
$\Gamma^A$, $A=1,\ldots 2N$, are some local coordinates). The density
should be compatible with the antibracket in such a way that the BV
$\Delta$ operator
\begin{equation}
  \Delta_\rho H=\half \div{\rho}(\vectorf V_{\!\!H})
  \label{BVDelta}
\end{equation}
be nilpotent, $\Delta_\rho^2=\commut{\Delta_\rho}{\Delta_\rho}=0$.
Here, $\div{\rho}$ denotes the divergence of a vector field with
respect to the density $\rho$ and $\vectorf
V_{\!\!H}=\abrkt{H}{\cdot~}$ is the globally Hamiltonian vector field
with the Hamiltonian~$H$.

The physics is determined by the quantum master-action
$W\in\func\cM[[\hbar]]$ (a formal power series in~$\hbar$ with
coefficients in~$\func\cM$) that satisfies the quantum master equation
\begin{equation}
  \Delta_\rho e^{\frac{i}{\hbar}W}=0\, \quad
  \Longleftrightarrow   \quad
  \half\abrkt{W}{W}=i\hbar\Delta_\rho W\,.
  \label{Qmaster}
\end{equation}
Writing $W=S+i \hbar W_1+(i \hbar)^2 W_2+\ldots$, we rewrite
\req{Qmaster} as
\begin{eqnarray}
  \abrkt{S}{S}&=&0\,, \label{Cmaster}\\
  \abrkt{S}{W_1}&=&\Delta_\rho S\,,
\end{eqnarray}
and so on. Equation \req{Cmaster} is the classical master equation,
and the function $S=W|_{\hbar=0}$ is called the classical master
action.

In addition to the master equation, one should impose boundary
conditions on~$W$. This requires fixing a Lagrangian submanifold
$\cL_0$ in $\cM$ (in the canonical coordinates, the manifold of
fields, with the antifields set to zero) and a function $\cS$ on
$\cL_0$, which is the {\it original (``bare'') action\/} of the
classical theory that is being quantised. Then one
requires~$W(\Gamma,\hbar)$ to be such that
$W(\cdot,0)|_{\cL_0}=\cS\,$.

By definition, a quantum observable is a function
$A\in\func\cM[[\hbar]]$ that satisfies $\delta_W A=0$, where
\begin{equation}
  \delta_W A=(W,A)-i \hbar\Delta_\rho{A}\,.
  \label{Qobserv}
\end{equation}
It follows from~\req{Qmaster} that $\delta_W^2=0$ and therefore any
function $A$ of the form $A=\delta_W B$ is an observable; these are
called {\it trivial\/} observables.  Expanding $A=A_0+i \hbar
A_1+(i\hbar)^2 A_2+\ldots$, we rewrite equation $\delta_W A=0$ as
\begin{eqnarray}
  \abrkt{S}{A_0}&=&0\,, \label{Cobserv}\\
  \abrkt{S}{A_1}+\abrkt{W_1}{A_0}&=&\Delta_\rho A_0\,,
\end{eqnarray}
and so on.  An $\hbar$-independent function~$A_0$
satisfying~\req{Cobserv} is called the {\it classical observable.\/}
It is easy to see that if $A=\delta_W B$ then $A_0=(S,B_0)$, where
$B_0=B|_{\hbar=0}$.  Any classical observable $A_0$ of the form
$A_0=(S,B_0)$ with some $\hbar$-independent function $B_0$ is called
the {\it trivial classical observable}.

The quantum expectation of an observable is defined via the
path-integral over a Lagrangian submanifold $\LG$,
\begin{equation}
  \langle A \rangle =
  \int\limits_\LG d\lambda_\rho A e^{\frac{i}{\hbar}W}\,,
  \label{PF}
\end{equation}
where $d\lambda_\rho$ is the volume form on~$\LG$ determined by the
volume form $d\mu=\rho d\Gamma$ on $\cM$ and by the antisymplectic
structure as follows~\cite{[ASS0],[BT]}:
\begin{equation}
  d\lambda_\rho(e^1,\ldots,e^N)=
  (d\mu(e^1,\ldots,e^N,f_1,\ldots,f_N))^{\frac{1}{2}} \,,
  \label{LGvolform}
\end{equation}
where $e^i \in T\LG$ and $f_j \in T\cM$ are any vectors that satisfy
$\twoform E(e^i\,,\,f_j)=\delta^i_j$ and $\twoform E$ is the
antisymplectic two-form on $\cM$.  It follows from \req{LGvolform}
that the volume form $d\lambda_{\rho^\prime}$ corresponding to the
density function $\rho^\prime=\rho e^H$ is related to $d \lambda_\rho$
as~$d\lambda_{\rho^\prime}=d\lambda_\rho e^{\half H}$ (this is the
origin of the exponent in Definition~\ref{def:Qsymm}).  If the
submanifold $\LG$ is determined by equations $G_\alpha=0$,
$\alpha=1,\ldots,N$, it would be Lagrangian whenever
$\abrkt{G_\alpha}{G_\beta}=U^\gamma_{\alpha\beta}G_\gamma \,$.

An important part of the BV axioms is the nondegeneracy conditions.
Submanifold $\LG$ in~\req{PF} must be such that the restriction of
$S=W|_{\hbar=0}$ to $\LG$ be nondegenerate.  In terms of the equations
$G_\alpha=0$, the matrix $\d_A G_\alpha$ and the Hessian matrix $\d_A
\d_B S$ should have no common null vectors at the points where
$\d_AS=0$ and $G_\alpha=0$~\cite{[BV],[BV2],[BT]}.  Whenever the set
$\nullm{\abrkt{S}{\cdot\,}}$ defined by equations $\d_AS=0$ is a
submanifold, this requirement means that $\nullm{\abrkt{S}{\cdot\,}}$
intersects $\LG$ transversely.  It also follows that the rank of the
Hessian matrix $H_{AB}=({\d_A \d_B S})|_{\d_A S=0}$ satisfies
$\rank{H_{AB}} \geq N$.  At the same time, the classical master
equation~\req{Cmaster} implies that $\rank{H_{AB}}\leq N$,
whence~\cite{[BV],[BV2],[BT]}
\begin{equation}
 \left. \rank{{\d^2 S\over\d \Gamma^A \d
      \Gamma^B}}\right|_{\d_A S=0}=N\,.  \label{proper}
\end{equation}
The solution of classical master-equation~\req{Cmaster} that
satisfies~\req{proper} is called a {\it proper solution}.

The key statement of the BV formalism is that the path integral
constructed as in~\req{PF} is invariant under infinitesimal
deformations of the Lagrangian
submanifold~$\LG$~\cite{[BV],[BV2],[ASS0],[BT]} for every quantum
observable $A$. In the case where $A=1$, this is often called the
gauge \hbox{independence of the partition function.}

\subsection{Lie algebras of gauge
  symmetries\label{sec:gauge-symmetries}} We now study Lie algebras of
gauge symmetries in the BV quantization scheme.  These are Lie
algebras $\cO^q$ and $\cO^c$ of {\it quantum and classical gauge
  symmetries}, respectively.  In addition to these two basic algebras,
it is useful to consider several more Lie algebras, which we define in
what follows and which can be arranged into the following commutative
diagram of homomorphisms of Lie algebras:
\begin{equation}\new
  \begin{array}{cccccc}
    \tilde\cO^q_{\rm triv}&\longrightarrow&\cO^q_{\rm triv}&
    \longrightarrow&\cO^q&\\
    \Bigm\downarrow&\kern-42pt\scriptscriptstyle{}^{\hbar=0}&
    \Bigm\downarrow&
    \kern-42pt\scriptscriptstyle{}^{\hbar=0}&\Bigm\downarrow&
    \kern-25pt\scriptscriptstyle{}^{\hbar=0}\\
    \tilde\cO^c_{\rm triv}&\longrightarrow&\cO^c_{\rm triv}&
    \longrightarrow&\cO^c&\\
    \begin{picture}(30,20)\unitlength=1pt
      \put(25,20){\vector(3,-2){45}}
    \end{picture}
    & &
    \begin{picture}(30,20)\unitlength=1pt
      \put(15,20){\vector(0,-2){27}}
    \end{picture}
    & &
    \begin{picture}(30,20)\unitlength=1pt
      \put(6,20){\vector(-3,-2){45}}
    \end{picture}
    &\\
    & &\kern+10pt\Ham_{\nullm\vect}& & &
  \end{array}\label{maindiag}\,
\end{equation}
In addition, we have homomorphisms~\req{exact-1}, \req{exact-2},
and~\req{exact-3}, whose constructions will also be explained in what
follows.  Here $\Ham_{\nullm\vect}$ is the Lie algebra of {\it
  on-shell gauge symmetries\/} (which, as we show, is the
  algebra of locally Hamiltonian vector fields on~$\nullm\vect$),
$\cO^q_{\rm triv}$ is the Lie algebra {\it of BRST-trivial quantum
  gauge symmetries\/}, and $\tilde\cO^q_{\rm triv}$ is the Lie algebra
{\it of quantum gauge parameters\/}~\cite{[SZ],[SZ2]}.  Lie algebras
$\cO^c$, $\cO^c_{\rm triv}$ and $\tilde\cO^c_{\rm triv}$ are the
classical counterparts of $\cO^q$, $\cO^q_{\rm triv}$, and
$\tilde\cO^q_{\rm triv}$ respectively.  We now proceed to the exact
definitions.
\begin{dfn}[\cite{[SZ],[ASS1]}]\label{def:Qsymm}\mbox{}
  A vector field $\vectorf X(\hbar)$ is called the quantum gauge
  symmetry if it preserves the antibracket and the measure $\rho
  e^{\frac{2i}{\hbar}W} d\Gamma$ (viewed as formal power series
  in~$\hbar$).

  The Lie algebra $\cO^q$ of these vector fields is called the Lie
  algebra of quantum gauge symmetries.
\end{dfn}
It follows from this definition that a quantum gauge
symmetry~$\vectorf X(\hbar)$ satisfies
\begin{equation}
  \div{\rho}(\vectorf X(\hbar))+\frac{2i}{\hbar}\vectorf X(\hbar)
  W=0\,, \label{QXpreserv}
\end{equation}
\begin{equation}
  \vectorf X\abrkt{F}{G}-
  \abrkt{\vectorf X F}{G}-
  (-1)^{(\e F+1)\e{\vectorf X}}\abrkt{F}{\vectorf X G}=0\,,
  \qquad F,G\in \func\cM\,.
  \label{ABRKTpreserv}
\end{equation}
Equation~\req{ABRKTpreserv} implies that there exists, at least
locally, a function $A(\hbar)$ such that $\vectorf
X(\hbar)=\abrkt{A(\hbar)}{\;\cdot\;}$.  Then Eq.~\req{QXpreserv}
implies that $(W,A(\hbar))-i\hbar\Delta_\rho{A(\hbar)}=0$.  Whenever
$A(\hbar)$ is globally defined, it is a quantum observable. We
explicitly indicate the $\hbar$-dependence of $\vectorf X (\hbar)$
because $\rho e^{\frac{2i}{\hbar}W}$ should be preserved for any value
of $\hbar$; we assume $\vectorf X (\hbar)$ to be a formal power series
in $\hbar$ with coefficients in the vector fields on $\cM$.

Although $\vectorf X(\hbar)$ is not a symmetry of any classical system
(in particular, it preserves neither the quantum master action nor the
measure $d \mu$), we call it a quantum gauge symmetry because its
classical counterpart, obtained by taking the limit as $\hbar \to 0$,
does preserve the classical master action~$S$. (The latter can be
considered as the action of some classical system defined on~$\cM$.
Then the classical master equation can be viewed as an additional
constraint imposed on the system with the action~$S$.  One can
naturally identify gauge symmetries of this system with the
transformations preserving both $S$ and the master equation
imposed~on~$S$.)

To make contact with the literature, we consider the Lie algebra
$\cO^q_{\rm triv}$ of quantum {\it BRST-trivial gauge symmetries\/}
studied in~\cite{[SZ]}.  These are quantum gauge symmetries $\vectorf
X_{B}(\hbar)=(\delta_W B(\hbar),\cdot~)$ whose Hamiltonians are
trivial observables (see~\req{Qobserv}), which span an ideal
in~$\cO^q$.

Now, the (Hamiltonian) mapping $\func\cM[[\hbar]]\to\cO^q_{\rm triv}$
allows us to pullback the Lie bracket from $\cO^q_{\rm triv}$ to the
space of $\hbar$-dependent functions.  Namely,
\begin{equation}
  \commut{B^1(\hbar)}{B^2(\hbar)}^q=
  \abrkt{B^1(\hbar)}{\delta_W B^2(\hbar)}\,,
  \label{Qbracket}
\end{equation}
which implies
\begin{equation}
  \commut{\vectorf X_{B^1}(\hbar)}{\vectorf X_{B^2}(\hbar)}
  =\abrkt{\delta_W \abrkt{B^1(\hbar)}{\delta_W B^2(\hbar)}}{\cdot~}
  =\vectorf X_{\commut{B^1(\hbar)}{B^2(\hbar)}^q}\,.
  \label{Qtrivial}
\end{equation}
The bracket \req{Qbracket} was shown in \cite{[SZ],[SZ2]} to determine
a Lie algebra structure on the quotient space
$$
{\tilde\cO}^q_{\rm
  triv}=\func\cM[[\hbar]]\Bigr/\delta_W\func\cM[[\hbar]]
$$
of all $\hbar$-dependent functions modulo the $\delta_W$-exact ones.
$\tilde\cO^q_{\rm triv}$ was called the {\it Lie algebra of quantum
  gauge parameters\/} in~\cite{[SZ],[SZ2]}.\footnote{This
  is not an algebra under the {\it associative\/} multiplication
  because the multiplication does not preserve the equivalence classes
  $B(\hbar) \sim B(\hbar)+\delta_W C(\hbar)$; in
  particular,~\req{Qbracket} is not a Poisson bracket.}

There is a nice way to `measure' how $\cO^q_{\rm triv}$ differs from
${\tilde\cO}^q_{\rm triv}$.  The (Hamiltonian) mapping
$\func\cM[[\hbar]]\to\cO^q_{\rm triv}$ induces a
homomorphism~$\tilde\cO^q_{\rm triv} \to \cO^q_{\rm triv}$ (see
diagram~\req{maindiag}), whose kernel consists of functions (modulo
$\delta_W$-exact ones) satisfying $\delta_W B(\hbar)={\rm
  const}(\hbar)$.  However, the fact that a function $F$ satisfies
$\delta_W F(\hbar)={\rm const}(\hbar)$ implies
$\delta_WF(\hbar)=0$.\footnote{In order to see this, consider first a
  function $F_0$ satisfying $\vect F_0= \abrkt{S}{F_0}={\rm const}$.
  Since (as we see in the next subsection) $\cM$ is a \pmanifold, the
  function $\vect F_0$ vanishes on the zero locus of~$\vect$ and
  therefore $\vect F_0=0$.
  Now, to see that equation $\delta_W F(\hbar)={\rm const}(\hbar)$
  leads to $\delta_W F(\hbar)=0$, we rewrite $\delta_W$ and $F(\hbar)$
  as power series in $\hbar$:
  $\delta_W=\delta^0_W+i\hbar\delta_W^1+(i\hbar)^2\delta_W^2+\ldots$,
  where in particular $\delta^0_W=\vect$, and $F=F_0+i\hbar F_1+(i
  \hbar)^2 F_2 +\ldots$.  Since $\cM$ is a \pmanifold, equation $\vect
  F_0=\abrkt{S}{F_0}=0$ implies that in some neighbourhood $F_0=\vect
  \phi_0+{\rm const}$ with some function $\phi_0$. Then in the first
  order in $\hbar$, the equation $\delta^1_W F_0+\delta^0_W F_1={\rm
    const}$ implies $\delta^1_W F_0+\delta^0_W F_1=0$ because
  $\delta^1_W F_0=-\delta_W^0 \delta_W^1 \phi_0$.  In higher orders in
  $\hbar$, a similar argument applies.  Thus the kernel of the
  homomorphism~$\tilde\cO^q_{\rm triv} \to \cO^q_{\rm triv}$ coincides
  with the cohomology of $\delta_W$ evaluated on the space of formal
  power series in $\hbar$ with the coefficients in smooth function on
  $\cM$.} We thus conclude that the homomorphism~$\tilde\cO^q_{\rm
  triv} \to \cO^q_{\rm triv}$ is included into the exact sequence that
involves the cohomology of $\delta_W$:
\begin{equation}\label{exact-1}
  0 \to \cH^q \to \tilde\cO^q_{\rm triv} \to \cO^q_{\rm triv} \to 0\,,
  \qquad \cH^q={\Ker\delta_W/\Im\delta_W}\,.\pagebreak[3]
\end{equation}

\medskip

The classical versions of these constructions are as follows.
\begin{dfn}\label{Csym} A vector field $\vectorf X_0$ is called a
  classical gauge symmetry if ${\vectorf X_0} S=0$ and $\vectorf X_0$
  preserves the antibracket.

  The Lie algebra $\cO^c$ of these vector fields is called the Lie
  algebra of classical gauge symmetries.
\end{dfn}

The classical {\it BRST-trivial gauge symmetries\/} are the vector
fields $\vectorf X_0=\abrkt{\abrkt{S}{B_0}}{\cdot~}$ whose
Hamiltonians are trivial classical observables (see~\req{Cobserv}).
These vector fields span the ideal $\cO^c_{\rm triv} \subset \cO^c$,
which is called the {\it classical BRST-trivial gauge symmetries.}

We have the obvious homomorphism~$\cO^q\stackrel{\scriptscriptstyle
  \hbar\to0}{\longrightarrow}\cO^c$.  This induces a homomorphism from
the ideal~$\cO^q_{\rm triv} \subset \cO^q$ into the ideal~$\cO^c_{\rm
  triv} \subset \cO^c$ (which are shown in~\req{maindiag}).

The classical counterpart of $\tilde\cO^q_{\rm triv}$ is the space
${\tilde\cO}^c_{\rm triv}$ of all functions on $\cM$ modulo the
functions of the form $\abrkt{S}{C}$, where $S$ is the classical
master action satisfying~$\abrkt{S}{S}=0$.  One can see that the space
${\tilde \cO}^c_{\rm triv}$ is endowed with a Lie algebra structure
with respect to the `classical' bracket
\begin{equation}
  \commut{B^1_0}{B^2_0}^c=\abrkt{B^1_0}{\abrkt{S}{B^2_0}}\,,
  \label{Cbracket}
\end{equation}
Thus we have the Lie algebra homomorphism~${\tilde\cO}^c_{\rm triv}
\to \cO^c_{\rm triv}$ shown in~\req{maindiag}.  The kernel of the
homomorphism coincides with the cohomology of $\vect$, therefore we
have the following exact sequence involving the cohomology of
$\vect{}$:
\begin{equation}\label{exact-2}
  0 \to \cH^c \to \tilde\cO^c_{\rm triv} \to \cO^c_{\rm triv} \to 0\,,
  \qquad \cH^c={\Ker\vect/\Im\vect}\,.
\end{equation}

We also observe that the relation between the quantum and the
classical bracket is given by $\commut{B^1_0}{B^2_0}^c=
\restrict{\commut{B^1(\hbar)}{B^2(\hbar)}^q}{\hbar=0}$, where
$B^i_0=B^i|_{\hbar=0}$. Therefore there exists a Lie algebra
homomorphism $\tilde\cO^q_{\rm triv}\stackrel{\scriptscriptstyle
  \hbar\to0}{\longrightarrow}\tilde\cO^c_{\rm triv}$.
Following~\cite{[SZ],[SZ2]} we call $\tilde\cO^c_{\rm triv}$ the {\it
  Lie algebra of classical gauge parameters}.  We thus see how it is
related to the other algebras in~\req{maindiag}.

Of the algebras entering~\req{maindiag}, it only remains to
construct~$\Ham_{\nullm\vect}$, which we now do in the BV-setting.

\subsection{The Hamiltonian algebra of on-shell gauge
  symmetries\label{subsec:Poisson-gauge}}

The BV field-antifield manifold $\cM$ and the classical master action
$S$ satisfying the BV quantization axioms are such that $\cM$ is a
\pmanifold.  Indeed, the odd vector field $\vect=\abrkt{S}{\cdot~}$ on
the $(N|N)$-dimensional antisymplectic manifold $\cM$ preserves the
antibracket and therefore satisfies condition~\req{Leibnitz}, the
master equation imposed on $S$ implies that $\vect$ is nilpotent, and,
finally, the fact that $S$ is a proper solution of the master equation
implies the rank condition~\req{rankcondit}.  The zero locus
$\nullm\vect$ of~$\vect$ determined by the equations $\d_AS=0$ will be
referred to as the {\it stationary surface\/} of the action~$S$.  As
before, we assume $\nullm\vect$ to be a smooth
submanifold\,\footnote{Although in realistic examples the structure of
  the zero locus of~$\vect$ can be very involved, we treat
  $\nullm\vect$ as a submanifold. Note in passing that the
  finite-dimensional models of gauge systems should be considered with
  some caution also in view of the results of~\cite{[BT-new]}.}.  Then
according to Theorem~\ref{thm:symplectic}, $\nullm\vect$ has a natural
symplectic structure. Further, the classical gauge symmetries (see
Definition~\ref{Csym}) are in fact symmetries of the
\pmanifold{}~$\cM$.
\begin{thm}
  Every classical gauge symmetry $\vectorf X_0$ determines a vector
  field $\vectorf x=\restrict{\vectorf X_0}{\nullm\vect}$ on
  $\nullm\vect$ that preserves the Poisson bracket on $\nullm\vect$.
\end{thm}
\begin{prf}
  Indeed, any vector field~$\vectorf X_0$ preserving the master action
  $S$ and the antibracket commutes with $\vect = \abrkt{S}{\cdot~}$
  and is therefore a symmetry of~$\cM$ (Definition~\ref{def:PQPsymm}).
  As we saw in section~\ref{QPsymm:subsec}, any vector field $\vectorf
  X_0$ that is a symmetry of~$\cM$ restricts to $\nullm\vect$ and
  $\restrict{\vectorf X_0}{\nullm\vect}$ is locally Hamiltonian on
  $\nullm\vect$.
\end{prf}

We denote the Lie algebra of locally Hamiltonian vector fields on
$\nullm\vect$ by $\Ham_{\nullm\vect}$.  As we are going to see, this
is the algebra of on-shell gauge symmetries.
\begin{dfn}
  A classical gauge symmetry $\vectorf X_0$ is called on-shell trivial
  if it vanishes on the stationary surface~$\nullm\vect$.
\end{dfn}
We now show that the Lie algebra of locally Hamiltonian vector fields
on $\nullm\vect$ is isomorphic to the algebra of on-shell gauge
symmetries.
\begin{thm}
  The algebra $\cI_0$ of on-shell trivial symmetries is an ideal in
  the Lie algebra~$\cO^c$ of gauge symmetries and the quotient algebra
  $\cO^c / \cI_0$ is isomorphic to the Lie algebra
  $\Ham_{\nullm\vect}$ of locally Hamiltonian vector fields
  on~$\nullm\vect$.
\end{thm}
\begin{prf}
  Let~$\vectorf Y_0 \in \cI_0$ and~$\vectorf X_0 \in \cO^c$. For any
  function~$F\in\func\cM$, we have~$\vectorf Y_0 F \in
  \ideal{\nullm\vect}$.  Since $\vectorf
  X_0\ideal{\nullm\vect}\subset\ideal{\nullm\vect}$ and $\vectorf Y_0
  F\in \ideal{\nullm\vect}$, we have \ $\commut{\vectorf X_0}{\vectorf
    Y_0}F= \vectorf X_0 \vectorf Y_0 F -(-1)^{\e{\vectorf X_0
      }\e{\vectorf Y_0 }} \vectorf Y_0 \vectorf X_0 F \in
  \ideal{\nullm\vect}$, therefore~$\cI_0$ is an ideal in the Lie
  algebra~$\cO^c$.  Further, we have seen in Theorem~\ref{thm:lift}
  that any locally Hamiltonian vector field $\vectorf x$ on
  $\nullm\vect$ is a restriction of some vector field $\vectorf X \in
  \cO^c$.  Thus one can identify the quotient algebra $\cO^c / \cI_0$
  with the Lie algebra of locally Hamiltonian vector
  fields~on~$\nullm\vect$.
\end{prf}

It follows from the theorem that the
homomorphism~$\cO^c\to\Ham_{\nullm\vect}$ is included into the exact
sequence
\begin{equation}\label{exact-3}
  0\to\cI_0\to\cO^c\to\Ham_{\nullm\vect}\to0\,.
\end{equation}

Note that we cannot replace $\cO^c$ with $\cO^c_{\rm triv}$ here,
because the homomorphism $\cO^c_{\rm triv}\to\Ham_{\nullm\vect}$ is
{\it not\/} surjective whenever there exists the first cohomology
of~$\nullm\vect$. Indeed, a nonvanishing first cohomology implies that
there exist locally Hamiltonian vector fields that are not globally
Hamiltonian on~$\nullm\vect$, which we have seen in
theorem~\ref{thm:lift} to correspond to {\it BRST-nontrivial\/} gauge
symmetries. Due to the existence of the latter, the mapping
$\cO^c_{\rm triv}\to\Ham_{\nullm\vect}$ is not surjective in general.

\smallskip

Looking at diagram~\req{maindiag}, it is natural to ask the following
question: What is the analogue of~$\Ham_{\nullm\vect}$ for the upper
line of the diagram, i.e.\ what is the {\it quantum\/} analogue of the
on-shell gauge symmetries?  We propose one possible answer to this
question.

Note that Poisson bracket~\req{Poissonbracket} on~$\nullm\vect$ and
Lie bracket~\req{Cbracket} on the space of gauge parameters are
defined by the same bilinear
operation~$\abrkt{~\cdot}{\vect\;\cdot\,}$ on~$\func\cM$.  The
difference between these two brackets is that~\req{Cbracket} is
defined on the quotient space~$\func\cM/\Im\vect$, while the Poisson
bracket is defined on~$\func\cM/\ideal{\nullm\vect}$. In the case of a
\pmanifold, $\ideal{\nullm\vect}$~is the ideal generated
by~$\Im\vect$, i.e.~$\ideal{\nullm\vect}=\func\cM\cdot\Im\vect$.  At
the same time,~\req{Cbracket} is the limit as~$\hbar\to 0$ of the
quantum construction~\req{Qbracket} defined on~$\func\cM[[\hbar]]$.
Therefore, in the quantum case one can construct a Poisson bracket as
a direct generalisation of~\req{Poissonbracket},
as~${\pbrkt\cdot\cdot}^q=\abrkt{\cdot}{\delta_W\,\cdot}$.  The
bracket~${\pbrkt\cdot\cdot}^q$ would be well defined only on the
quotient algebra of~$\func\cM[[\hbar]]$ modulo the
ideal~$\ideal{\delta_W}$ generated by~$\Im\delta_W$.
Obviously,~$\Im\delta_W$ is generated by all series of the
form~$\delta_W f(\hbar)$,
where~$f(\hbar)=f_0+f_1\hbar+f_2\hbar^2+\ldots$ with the
coefficients~$f_i$ taking independently each value~$1$, $\Gamma^A$
and~$\Gamma^A\Gamma^B$.  Since the
matrix~$E^{AB}=\abrkt{\Gamma^A}{\Gamma^B}$ is invertible, we thus see
that~$\ideal{\delta_W}$ consists of the series of the
form~$\ideal{\nullm\vect}+\func\cM\hbar+\func\cM\hbar^2+\ldots$ .
Thus, the quotient algebra $\func\cM[[\hbar]]/\ideal{\delta_W}$
coincides with the algebra~$\func{\nullm\vect}$ of functions on the
zero locus of~$\vect$.  This means that the
algebra~$\Ham_{\nullm\vect}$ is in a certain sense the most general
algebra of the on-shell symmetries not only of the classical but also
of the quantum, master action.

\section{Examples\label{Examples}}
\subsection{Abelianized gauge theory}
We now consider the field-antifield space and the master action
corresponding to the simplest gauge theory, the abelianized gauge
theory, which we choose as an instructive example that is free of
additional complications because gauge symmetries are explicitly
separated from the physical ones.  We then explicitly construct the
Poisson bracket and the Lie algebras $\cO^c$ and $\Ham_{\nullm\vect}$.
Moreover, this example shows that the classical gauge symmetries
$\cO^c$ contain the Lie algebra of gauge symmetries of the {\it
  original\/} theory as a subalgebra and that, similarly, the on-shell
gauge symmetries $\Ham_{\nullm\vect}$ contain the on-shell symmetries
of the abelianized gauge theory.

Let $S_0(X,x)$ be a polynomial action such that
\begin{equation} \label{rnk}
  \d_\alpha S_0=0\,,\quad
  \restrict{\det\left(\d_i\d_jS_0\right)}{\d_iS_0=0}\neq0\,,
\end{equation}
where we denote $\d_\alpha=\frac{\d}{\d x^\alpha}$ and
$\d_i=\frac{\d}{\d X^i}$ and assume $X^i$ and $x^\alpha$ to be bosonic
for simplicity.  Due to rank condition~\req{rnk}, the equations~$\d_i
S_0=0$ admit only a finite set of solutions~$\M$.  Thus, the
stationary surface of this theory is the direct product of~$\M$ with
the space parametrised by~$x^\alpha$.  The gauge transformations
preserving the action~$S_0$ are of the form
\begin{equation}\label{4.2}
  \vectorf Y_0=Y_0^\alpha(X,x)\d_\alpha + \mu^{ij}(X,x)\d_iS_0\d_j\,,
\end{equation}
where $\mu^{ij}(X,x)$ is an antisymmetric matrix.  These vector fields
span a Lie algebra~$\cA$ with respect to the commutator of vector
fields; those vanishing on the stationary surface span the
ideal~$\cA_{\rm triv}$ in~$\cA$.  Then~$\tilde\cA=\cA/\cA_{\rm triv}$
is the algebra of on-shell gauge symmetries, which can be identified
with the Lie algebra of vector fields on the stationary surface.

To carry out the BV scheme, we choose the gauge generators in the
form~$R^\alpha_\beta=\delta^\alpha_\beta$ and introduce ghosts
$c^\alpha$ and the antifields $\ax_\alpha$, $X^*_i$, and~$c^*_\alpha$.
The canonical antibracket is $\abrkt{\phi^A}{\phi^*_B}=\delta^A_B$,
where $\phi^A=(X^i$, $x^\alpha,c^\alpha)$ and
$\phi^*_A=(X^*_i$,$\ax_\alpha,c^*_\alpha)$.  Then the master action
\begin{equation}
  S=S_0+\ax_\alpha c^\alpha
  \label{abact}
\end{equation}
is a proper solution of the master equation~$\abrkt S S=0$.  This
action defines the vector field
\begin{equation}
  \vect=\abrkt{S}{\cdot}=\d_i S_0\dr{X^*_i}+c^\alpha\dr{x^\alpha}
  +x^*_\alpha\dr{c^*_\alpha}\,,
\end{equation}
whose stationary surface $\nullm\vect$ is determined by $\d_i S_0=0$,
$\ax_\alpha=0$, $c^\alpha=0$.  Thus~$\nullm\vect$ is the direct
product of $\M$ (the set of solutions to the system of equations~$\d_i
S_0=0$) with the space parametrised
by~$Y^A=\{X^*_i,x^\alpha,c^*_\alpha\}$. The Poisson
bracket~\req{Poissonbracket} on~$\nullm\vect$ is then represented by
the matrix
\begin{equation}
  \Omega^{AB}=\pbrkt{Y^A}{Y^B}=\left(
    \begin{array}{ccc}
      \d_i\d_jS_0  & 0 & 0 \\
      0  & 0 & \delta_\alpha^\beta \\
      0  & -\delta_\gamma^\nu & 0
    \end{array}\right)\,.
\end{equation}

We now want to show that the Lie algebras~$\cA$ and~$\tilde\cA$ of the
original theory are subalgebras in the Lie algebras~$\cO^c$
and~$\Ham_{\nullm\vect}$, respectively.  To do so, we
calculate~$\cO^c$ in the master theory with the master
action~\req{abact}.  Note that in the case of the abelianized gauge
theory, the first cohomology group of the field-antifield space
vanishes, therefore each Hamiltonian vector field has a globally
defined Hamiltonian.  Thus in order to find~$\cO^c$, it suffices to
find the kernel of~$\vect$ evaluated on the space of globally defined
functions.  Any element~$A\in\Ker\vect$ can be written in the form
$A=\vect F+G$, where $F$ is an arbitrary smooth function on the
field-antifield space and $G$ is a representative of the cohomology
class of~$\vect$.  In order to calculate the cohomology
of~$\vect$\,\footnote{ Note that in the case where
  $S_0=\half\delta_{ij}X^iX^j$, the vector field~$\vect$ is nothing
  but the de Rham differential of~$\nullm\vect$.  In this case the
  cohomology of~$\vect$ consists of constants only and~$\cO^c$
  coincides with~$\cO^c_{\rm triv}$.  }, we
write~$\vect=\vect_1+\vect_2$, where
\begin{equation}
  \vect_1=\d_iS_0\dr{X^*_i}\,,\qquad
  \vect_2=c^\alpha\dr{x^\alpha}+x^*_\alpha\dr{c^*_\alpha}
  \quad\mbox{\rm and}\quad\vect_1^2=\vect_2^2=
  \commut{\vect_1}{\vect_2}=0\,.
\end{equation}

By the Poincar\'e lemma, the cohomology of $\vect_2$ consists of
constants only.  Thus the cohomology of~$\vect$ is determined by the
cohomology of~$\vect_1$ on the space of functions~$F(X,X^*)$.  A
function~$F(X,X^*)$ belongs to the image of~$\vect_1$
whenever~$F(X,X^*)=\d_iS_0 f^i(X,X^*)$, i.e.~$F(X,X^*)$ vanishes at
each point where~$\d_iS_0=0$.  Thus, any element~$A$ from~$\Ker\vect$
is of the form
\begin{equation}
  A(X,X^*) = \vect F(X,X^*) + G(X)\,,
\end{equation}
where $F$ is an arbitrary function and $G(X)$ is a function that
does not vanish at least at one point from~$\M$.  Whenever $\M$ is an
$n$-point set, the cohomology of $\vect$ is an $n$-dimensional vector
space\footnote{In this example, the group of `physical' symmetries is
  the group of the permutations of these $n$ points.  This group
  obviously acts on the cohomology of~$\vect$.}.

We thus see that $\cO^c$ and $\cO^c_{\rm triv}$ are spanned by the
vector fields of the form~$\abrkt{\vect F+G(X)}{\cdot~}$
and~$\abrkt{\vect F}{\cdot~}$ respectively. The algebra
$\Ham_{\nullm\vect}$ of on-shell symmetries consists of Hamiltonian
vector fields~$\pbrkt{H(\m,X^*,x,c^*)}{\cdot~}$ on~$\nullm\vect$
(where we label the Hamiltonian by $\m\in\M$ enumerating the different
components of~$\M$).

To see that the algebra~$\cA$ of gauge transformations of the original
theory is embedded into the Lie algebra~$\cO^c$ of the classical gauge
symmetries, we note that vector fields of the form
\begin{equation}\label{4.12}
  \vectorf Y=
  \abrkt{Y_0^\alpha(X,x)x^*_\alpha + \mu^{ij}(X,x)\d_iS_0X^*_j}{\cdot~}
\end{equation}
form the subalgebra in $\cO^c$. Moreover, these fields restrict to the
subspace $c^\alpha=c^*_\alpha=x^*_\alpha=X^*_i=0$ as elements
of~$\cA$ (see~\req{4.2}).  Thus we have an embedding of $\cA$ into
$\cO^c$ (obviously, the embedding is not unique).

As regards the on-shell gauge symmetries, observe that vector fields
on~$\nullm\vect$ of the form
\begin{equation}
  \vectorf y=\pbrkt{y^\alpha(\m,x)c^*_\alpha}{\cdot~}
\end{equation}
(which define a subalgebra in~$\Ham_{\nullm\vect}$) restrict to the
stationary surface of the original theory (which is a submanifold of
~$\nullm\vect$ determined by the equations~$c^*_\alpha=0$
and~$X^*_i=0$) and span~$\tilde\cA$.  Thus the algebra of on-shell
gauge symmetries~$\tilde\cA$ of the original theory is embedded into
the Lie algebra~$\Ham_{\nullm\vect}$ of the on-shell gauge symmetries.

\subsection{A `topological' field theory}
We now apply Theorem \ref{thm:symplectic} to the `topological' theory
with the vanishing action on a Lie group~$\cG$.  We show that in this
case bracket~\req{Poissonbracket} is related to the Kirillov bracket
on the coalgebra.  Denote by~$x^i$ a coordinate system in the
neighbourhood of~$1\in\cG$.  Let~$\cR_\alpha=R^i_\alpha\d_i$ (where
the Greek indices have the same cardinalities as the Latin ones) be
the basis of the left invariant vector fields on~$\cG$. We have
$\commut{\cR_\alpha}{\cR_\beta}=\cF^\gamma_{\alpha\beta}\cR_\gamma$,
where $\cF^\gamma_{\alpha\beta}$ are the structure constants.

In accordance with the BV prescription, we introduce the ghosts
$c^\alpha$ and the antifields $x^*_i$ and $c^*_\alpha$ such that
$\abrkt{x^i}{x^*_j}=\delta^i_j$, $\abrkt{c^\alpha}{c^*_\beta}
=\delta^\alpha_\beta$. The master action \ $S=x^*_iR_\alpha^ic^\alpha
-\half c^*_\gamma\cF^\gamma_{\alpha\beta}c^\beta c^\alpha$ \ is a
proper solution of~$\abrkt{S}{S}=0$.  Then
\begin{equation}
  \vect=\abrkt{S}{\cdot~}= c^\alpha R_\alpha^i\dr{x^i}
  +\half\cF^\gamma_{\alpha\beta}c^\beta c^\alpha \dr{c^\gamma}
  +(x^*_i R_\alpha^i - c^*_\gamma\cF^\gamma_{\alpha\beta}c^\beta)
  \dr{c^*_\alpha}\,,
\end{equation}
therefore the zero locus~$\nullm\vect$ of~$\vect$ is determined by the
equations \ $c^\alpha=0$, $x^*_i=0$ \ and is coordinatized
by~$Y^A=\{x^i\,,\,c^*_\alpha\}$.  In this
case~$\nullm\vect=T^*\cG=\cG\times\g^*$ is the cotangent bundle
to~$\cG$, where $\g^*$ is the coalgebra.  The matrix of Poisson
bracket~\req{Poissonbracket} takes the form
\begin{equation}
  \Omega^{AB}=\pbrkt{Y^A}{Y^B}=
  \left(\begin{array}{cc}
      \pbrkt{x^i}{x^j}            & \pbrkt{x^i}{c^*_\beta}\\
      \pbrkt{c^*_\alpha}{x^j}     & \pbrkt{c^*_\alpha}{c^*_\beta}
    \end{array}\right)=
  \left(\begin{array}{cc}
      0     & R^i_\beta\\
      -R^j_\alpha     & -c^*_\gamma\cF^\gamma_{\alpha\beta}
    \end{array}\right)\label{Kiril}\,.
\end{equation}
It is nondegenerate because~$R_\alpha^i$ are nondegenerate everywhere
on~$\nullm\vect$ since~$R_\alpha^i$ are the coefficients of the left
invariant vector fields on the Lie group. The cotangent bundle to a
Lie group is trivial, therefore we have the embedding~$\g^*\rightarrow
T^*\cG$, which induces a Poisson bracket on~$\g^*$.  This gives us the
Kirillov bracket~\cite{[Kir]} on the coalgebra~$\g^*$ parametrised by
the coordinates~$c^*$.

\section{Conclusions}
We have seen that a number of objects of the antisymplectic BV
geometry are essentially determined by objects of the symplectic
geometry on the stationary surface of the master action, where the
nondegenerate Poisson bracket is given by~\req{Poissonbracket}.
In particular, every observable determines a symmetry of the
master-action, which in turn restricts to a locally Hamiltonian vector
field on $\nullm\vect$; at the same time, every trivial observable
determines a symmetry of the master action such that the corresponding
vector field on $\nullm\vect$ is globally Hamiltonian.
Those Hamiltonian vector fields on~$\nullm\vect$ that are not globally
Hamiltonian correspond then to the BRST-nontrivial observables.

Recalling how the master theory is constructed in terms of the bare
classical action~$\cS$, we were able to explicitly see, in the
abelianized setting, that the gauge symmetries of~$\cS$ are dressed
into symmetries of the master theory, i.e., into $(~,\;)$-Hamiltonian
vector fields; at the same time, on-shell gauge symmetries of~$\cS$
are dressed into $\{~,\;\}$-Hamiltonian vector fields on the
symplectic manifold~$\nullm\vect$. This would be interesting to extend
the setting of the general gauge theory.

Our analysis was performed in the framework of the finite-dimensional
model; such models should be viewed with caution precisely for the
reasons related to the existence of the BRST
cohomology~\cite{[BT-new]}. It would be interesting to see how our
results can be reformulated in local field theory, where the gauge
symmetries have been discussed in~\cite{[Henn]}, and, possibly, also
in application to string field theory~\cite{[HZ],[SZ2]}, which has
been one of the motivations behind the geometrically covariant
reformulation of the BV quantization.

\paragraph{Acknowledgements.} We are grateful to I.~Batalin and
I.~Tyutin for many illuminating discussions on various aspects of the
field-antifield quantization and the related problems. AMS wishes to
thank P.H.~Damgaard for a helpful discussion and for kind hospitality
at the Niels Bohr Institute.  We also appreciated discussions with
O.~Khudaverdyan, A.~Nersessian, and B.~Voronov.  This work was
supported in part by the RFBR Grant~96-01-00482 and by the INTAS-RFBR
Grant~95-0829.


\small


\begin{thebibliography}{33}
  \parindent=0pt \parskip=-2pt

\bibitem{[BV]} I.~A.~Batalin and G.~A.~Vilkovisky, \PLB{102} (1981)
  27\,,

\bibitem{[BV2]} I.~A.~Batalin and G.~A.~Vilkovisky, \PRD{28} (1983)
  2567.

\bibitem{[BT]} I.~A.~Batalin and I.~V.~Tyutin, Int. J.  Mod.  Phys. A8
  (1993) 2333;
  Mod. Phys.  Lett.~A8~(1993)~3673;
  Mod. Phys.  Lett.~A9~(1994)~1707.

\bibitem{[ASS0]} A.~Schwarz, \CMP{155}~(1993)~249--260.

\bibitem{[HZ]} H.~Hata and B.~Zwiebach, Ann. Phys. 229 (1994)
  177--216,

\bibitem{[SZ2]} A.~Sen and B.~Zwiebach, \CMP{177} (1996) 305.

\bibitem{[ASS3]} A.~Schwarz, \CMP{158} (1993) 373--396.

\bibitem{[KN]} O.~M.~Khudaverdian, J. Math. Phys. 32 (1991) 1934;
  O.~M.~Khudaverdian and A.~Nersessian, \MPLA8 (1993) 2377.

\bibitem{[Bering]} K.~Bering, {\it Almost Parity Structure,
    Connections and Vielbeins in BV Geometry}, MIT-CTP-2682,
  physics/9711010.

\bibitem{[SZ]} A.~Sen and B.~Zwiebach, \PLB{320} (1994) 29--35.

\bibitem{[ASS2]} M.~Alexandrov, M.~Kontsevich, A.~Schwarz and
  O.~Zaboronsky, \IJMPA{12} (1997) 1405--1430.

\bibitem{[DN]} A. Nersessian and P.H. Damgaard,
  \PLB355 (1995) 150.

\bibitem{[BMS]} I.~A.~Batalin, R.~Marnelius, and A.~M.~Semikhatov,
  \NPB{446} (1995) 249.

\bibitem{[ASS1]} A.~Schwarz, Lett. Math. Phys. 31 (1994) 299--302.

\bibitem{[Ners]} O.~M.~Khudaverdian and A.~P.~Nersessian, \JMP{37}
  (1996) 3713--3724.
  A.~P.~Nersessian\,, Preprint JINR E2-93-225;
  Preprint JINR E2-93-358.



\bibitem{[Kir]} A.~A.~Kirillov, ``Elements of the theory of
  representations'' Springer, Berlin, 1970.

\bibitem{[BT-new]} I.~A.~Batalin and I.~V.~Tyutin, Theor. Math.  Phys.
  114 (1998) 198-214.

\bibitem{[B-abel]} I.~A.~Batalin and G.~A.~Vilkovisky,
  Nucl. Phys. B234(1984) 106--124;\\
  I.~A.~Batalin and E.~S.~Fradkin,
  J. Math. Phys. 25 (1984) 2426--2429;\\
  B.~L.~Voronov and I.~V.~Tyutin, Teor. Mat. Fiz. 50 (1982) 333.

\bibitem{[Henn]} G.~Barnich and M.~Henneaux, \JMP{37} (1996)
  5273--5296.

\end{thebibliography}
\end{document}